\documentclass{appolb}
\usepackage{epsfig}
\usepackage{amsmath}
\usepackage{axodraw}
\usepackage{amssymb}
\usepackage{mathrsfs}
\usepackage{cite}

\begin{document}
\title{ENERGY CONSERVATION and POMERON LOOPS\\
in HIGH ENERGY EVOLUTION%
\thanks{Presented at the XLVI Cracow School of Theoretical Physics,
May 27 - June 5 2006, Zakopane, Poland.}%
}
\author{EMIL AVSAR
\address{Dept. of Theoretical Physics, 
S\"olvegatan 14A, S-223 62 Lund, Sweden\\
emil@thep.lu.se}}
\maketitle
\begin{abstract}
We present a formalism which modifies the Mueller Dipole Model 
such that it incorporates energy-momentum conservation and also 
important colour suppressed effects. We implement our formalism in a 
Monte Carlo simulation and compare the results to inclusive data from HERA and the
Tevatron, where we see that there is a good agreement between 
the data and our model.

\end{abstract}
  
\section{Introduction}

A convenient approach to small-$x$ evolution in QCD is 
the dipole picture which was formulated by Mueller a decade 
ago \cite{Mueller:1993rr,Mueller:1994jq,Mueller:1994gb}. Mueller's 
formalism leads to the BFKL equation but it also goes beyond
the BFKL formalism, since here it is possible to take 
into account unitarisation effects due to multiple scatterings
(multiple pomeron exchanges). In Mueller's model  
dipoles in the same cascade (wavefunction) are, however, not allowed to
interact and the model is therefore valid only in a limited range in
rapidity, where this type of saturation effects can be ignored. 
Saturation effects are included in the so called Colour Glass 
Condensate (CGC) formalism \cite{Iancu:2002xk,Iancu:2003xm}, 
which is an effective theory for QCD valid at high gluon densities. 
The master equation within the CGC formalism is called the 
JIMWLK equation which in the weak field limit reduces to the
BFKL equation. 

In this paper we continue to develop our model, introduced 
in \cite{Avsar:2005iz}, by adding $N_c$ suppressed 
effects in the dipole evolution. We also improve our 
proton model and use it to study inclusive data from 
$\gamma^*p$ and $pp$ collisions.   

\section{The Mueller Dipole Model}

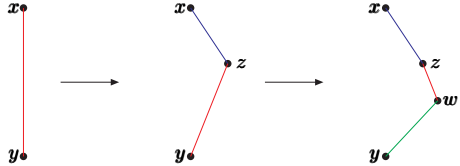
\begin{figure}
\begin{center}
\scalebox{0.7}{\mbox{
\begin{picture}(250,80)(0,5)
\Vertex(10,80){2}
\Vertex(10,0){2}
\Text(5,80)[]{$\pmb{x}$}
\Text(5,0)[]{$\pmb{y}$}
\SetColor{Red}
\Line(10,80)(10,0)
\SetColor{Black}
\LongArrow(30,40)(60,40)
\Vertex(100,80){2}
\Vertex(100,0){2}
\Vertex(120,50){2}
\Text(95,80)[]{$\pmb{x}$}
\Text(95,0)[]{$\pmb{y}$}
\Text(128,50)[]{$\pmb{z}$}
\SetColor{Blue}
\Line(100,80)(120,50)
\SetColor{Red}
\Line(120,50)(100,0)
\SetColor{Black}
\LongArrow(140,40)(170,40)
\Vertex(205,80){2}
\Vertex(225,50){2}
\Vertex(233,30){2}
\Vertex(205,0){2}
\Text(200,80)[]{$\pmb{x}$}
\Text(200,0)[]{$\pmb{y}$}
\Text(233,50)[]{$\pmb{z}$}
\Text(241,30)[]{$\pmb{w}$}
\SetColor{Blue}
\Line(205,80)(225,50)
\SetColor{Red}
\Line(225,50)(233,30)
\SetColor{Green}
\Line(233,30)(205,0)
\end{picture}
}}
\end{center}
\caption{\label{dipev} The evolution of the dipole cascade. 
At each step, a dipole can split into two new dipoles 
with decay probability given by~\eqref{eq:dipkernel}. }
\end{figure}

In Mueller's model we start with a $q\bar{q}$ pair, heavy enough
for perturbative calculations to be applicable, and calculate
the probability to emit a soft gluon from this pair. Here the 
quark and the antiquark are assumed to follow light-cone trajectories
and the emission of the gluon is calculated in the eikonal 
approximation, in which the emitters do not suffer any recoil. Adding the 
contributions to the emission from the quark and the antiquark,
including the interference, 
one obtains the result (for notations, see figure \ref{dipev}) 
\begin{eqnarray}
\frac{d\mathcal{P}}{dY}=\frac{\bar{\alpha}}{2\pi}d^2\pmb{z}
\frac{(\pmb{x}-\pmb{y})^2}{(\pmb{x}-\pmb{z})^2 (\pmb{z}-\pmb{y})^2}
\equiv \frac{\bar{\alpha}}{2\pi}d^2\pmb{z} \mathcal{M}(\pmb{x},\pmb{y},\pmb{z}).
\label{eq:dipkernel}
\end{eqnarray}
Here $\pmb{x}$, $\pmb{y}$, and $\pmb{z}$ are two-dimensional vectors
in transverse coordinate space and $Y=$log$(1/x)$ denotes the rapidity, which 
acts as the time variable in the evolution process.

This formula can be interpreted as the emission probability from a dipole
located at $(\pmb{x},\pmb{y})$. In the large $N_c$ limit the gluon 
can be seen as a quark-antiquark pair and the formula above can 
then be interpreted as the decay of the original dipole $(\pmb{x},\pmb{y})$
into two new dipoles, $(\pmb{x},\pmb{z})$ and $(\pmb{z},\pmb{y})$. 
In the same limit further emissions factorize, so that at each step
one has a chain of dipoles where each dipole can decay into 
two new dipoles with the decay probability given by~\eqref{eq:dipkernel}.
In this way one obtains a cascade of dipoles which evolve through
dipole splittings, and the number of dipoles grows exponentially with $Y$.

\section{Energy Momentum Conservation}

We note that the expression above has non-integrable singularities 
at $\pmb{z}=\pmb{x}$ and $\pmb{z}=\pmb{y}$. In numerical calculations 
it is therefore necessary to introduce
a cutoff, $\rho$, such that $(\pmb{x}-\pmb{z})^2,(\pmb{z}-\pmb{y})^2 \geq \rho^2$.
Even though this cutoff does not
show up in the cross section (the divergence is canceled 
by virtual corrections, and $\sigma_{\mathrm{tot}}$ approaches a constant 
when $\rho \rightarrow 0$) it must still be kept in a Monte Carlo program. 
A small value of $\rho$, which is needed in order to simulate 
the physics with a good accuracy, will imply that we get very many
small dipoles in the cascade. 

A small dipole means that 
we have two well localized gluons in the transverse plane, 
and these gluons must then have a correspondingly large 
transverse momentum of the order of the inverse dipole size,
$p_\perp \sim 1/r$. Thus if these small dipoles are 
interpreted as corresponding to real emissions with $p_\perp \sim 1/r$,
then the diverging number of such dipoles would imply the
violation of energy-momentum conservation. 
This suggest that these dipoles should be interpreted as virtual 
fluctuations, which means that the dipole cascade 
will not correspond to the production of exclusive final 
states.

In \cite{Avsar:2005iz} we presented a formalism to take into 
account energy-momentum conservation in Mueller's model, 
using the Linked Dipole Chain (LDC) model \cite{Andersson:1995ju} 
as a guidance. We used this formalism in a MC program and 
it was found that a result of energy-momentum
conservation is that the number of dipoles grows much more slowly, and 
the onset of saturation is delayed. In fact
it is found that in DIS the unitarity effects become quite 
small within the HERA energy regime. 

An important consequence of energy-momentum conservation 
is that it implies a \emph{dynamical cutoff}, $\rho(\Delta y)$, 
which is large for small steps in rapidity, $\Delta y$, but 
gets smaller for larger $\Delta y$. (Alternatively it can be 
described as a cutoff for $\Delta y$ which depends on $r$. 
Here we want to emphasize that $y$ is the true rapidity 
and not log$(1/x)$.) Besides its physical effects, energy-momentum 
conservation also simplifies the MC treatment, since large numerical
complications in a MC without energy conservation, as 
discussed in \cite{Salam:1995uy}, are not present. 

\section{Evolution Equations in High Energy QCD and Pomeron Loops}

We now consider a scattering process where we have a dipole 
(or a collection of dipoles) impinging on some arbitrary 
target. We denote the scattering amplitude for a single dipole 
by $\langle T\rangle$, and the scattering amplitude for $k$ dipoles
is for simplicity denoted by $\langle T^k\rangle$. The brackets 
here denote an averaging over different events. The high energy 
evolution equations for these amplitudes are determined by the so 
called Balitsky-JIMWLK (B-JIMWLK) equations 
\cite{Balitsky:1995ub,Jalilian-Marian:1997jx,Jalilian-Marian:1997gr,Iancu:2001ad,Weigert:2000gi}. In the large $N_c$ limit, the equation for $\langle T^k\rangle$
can be written as 
\begin{eqnarray}
\partial_Y \langle T^k\rangle = \mathcal{K}\otimes \langle T^k\rangle
- \mathcal{M}\otimes \langle T^{k+1}\rangle.
\label{eq:bjimwlk1}
\end{eqnarray}
Here we see that we have an infinite set of coupled equations 
for the amplitudes. We also note that $\langle T^k\rangle$ gets 
contributions from all $\langle T^n\rangle$ with $n\geq k$. 

If the rapidity increment is given to the projectile, the appearance
of $\langle T^{k+1}\rangle$ in~\eqref{eq:bjimwlk1} simply means
that one of the $k$ dipoles can split into two new dipoles with 
probability density given by~\eqref{eq:dipkernel}, and that both of
these scatter off the target, so that there are in total $k+1$ 
dipoles interacting with the target. This term therefore corresponds 
to pomeron splitting, since there is an exchange of $k+1$ pomerons 
instead of $k$ pomerons. Thus dipole splittings in the projectile 
corresponds to pomeron splittings. On the other hand one can think that 
the target is evolved rather than the projectile. In this case the 
$\langle T^{k+1}\rangle$ contribution to~\eqref{eq:bjimwlk1} corresponds 
to pomeron merging inside the target, since there are only $k$ dipoles 
in the projectile which couple to $k$ pomerons. 

We thus see that the B-JIMWLK equations describe either
pomeron mergings, when the target is evolved, or
pomeron splittings, in case the projectile is 
evolved, but not both. Ever since it was realized that the B-JIMWLK equations
are not complete, there has been a lot of effort to construct a
model which contains both pomeron mergings and splittings, and,  
through iterations, pomeron loops. This has been done in 
the large $N_c$ limit \cite{Iancu:2004iy,Mueller:2005ut} where the dipole
model has been used to add pomeron splittings to the B-JIMWLK equations
in the dilute region. The extension to the dense region 
is then obtained by simply adding the remaining 
terms arising from the large $N_c$ version of the B-JIMWLK
hierarchy. The main principle is that the two kinds of 
pomeron interactions (splittings and mergings) are important
in different, well separated, kinematical regions. 

The new equation for $\langle T^k\rangle$ receives a contribution
also from $\langle T^{k-1}\rangle$ and it can be written as 
\begin{eqnarray}
\partial_Y \langle T^k\rangle = \mathcal{K}\otimes \langle T^k\rangle
- \mathcal{M}\otimes \langle T^{k+1}\rangle + \mathscr{F}\otimes \langle T^{k-1}\rangle
\label{eq:bjimwlk2}
\end{eqnarray}
where $\mathscr{F}$ is a quite complicated expression
describing the fluctuations in the target (or saturation
effects in the projectile). 

\section{Finite $N_c$ Effects in Dipole Language}

In the dipole model saturation effects are included in the 
collisions between two cascades, but not in the evolution
of each cascade separately. 
Multiple collisions are formally colour suppressed, and in the Lorentz frame 
where the collisions are studied they lead to the formation of pomeron loops. 
As the evolution is only leading in $N_c$ such loops cannot be formed 
during the evolution itself.  If one e.g. studies the collision in a very 
asymmetric frame, where one of the onia carries almost all the available 
energy and the other is almost at rest, 
then the possibility to have multiple collisions is strongly reduced.
Only those pomeron loops are included, which are cut in the specific 
Lorentz frame used for the calculations, which obviously only forms a limited 
set of all possible pomeron loops. It implies that the dipole model is not 
frame independent, and the preferred Lorentz frame is the one where the two 
colliding systems have approximately the same density of dipoles. 
It is apparent that a frame independent formulation must include 
colour suppressed interactions between the dipoles during the evolution of 
the cascades, but so far it has not been possible to formulate such a model, 
which includes saturation effects and is explicitely frame independent
\footnote{Except for the toy model in which transverse dimensions 
are neglected.}.

As described above, equation~\eqref{eq:bjimwlk1} can be 
understood in terms of dipole evolution. The kernel of the evolution
is simply given by the dipole kernel $\mathcal{M}$ in~\eqref{eq:dipkernel}. 
Since pomeron splittings can be interpreted as dipole splittings
it has been hoped that pomeron mergins also can be interpreted 
in terms of dipole reactions, such as dipole mergings, \ie that 
one can interpret equation~\eqref{eq:bjimwlk2} in terms of a system
of dipoles which can either split or merge. However, it was shown in
\cite{Iancu:2005dx} that this is not quite possible. Formally this can be done, 
but the problem is that the would-be dipole merging vertex has no fixed sign, 
as would have been required by a proper probabilistic formalism. Before going 
on we note that one can generate pomeron mergings without the process 
of dipole mergings. For example, any $2\rightarrow n$ dipole vertex, 
where $n \geq 2$, also leads to pomeron mergings since there is 
always the possibility that only one of the new dipoles 
interact with the target, while the rest are spectators. Of course 
such a transition leads to all possible $2 \rightarrow m$ 
($m = 1, \dots,n$) pomeron transitions.          

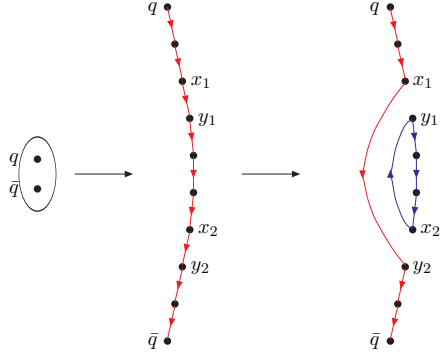
\begin{figure}
\begin{center}
\scalebox{0.7}{\mbox{
\begin{picture}(250,200)(0,5)
\Oval(10,110)(20,10)(0)
\Vertex(10,118){2}
\Text(0,118)[r]{$q$}
\Vertex(10,102){2}
\Text(0,102)[r]{$\bar{q}$}
\LongArrow(30,110)(60,110)
\Vertex(80,200){2}
\Text(75,200)[r]{$q$}
\Vertex(84,180){2}
\Vertex(88,160){2}
\Text(93,160)[l]{$x_1$}
\Vertex(92,140){2}
\Text(97,140)[l]{$y_1$}
\Vertex(94,120){2}
\Vertex(94,100){2}
\Vertex(92,80){2}
\Text(97,80)[l]{$x_2$}
\Vertex(88,60){2}
\Text(93,60)[l]{$y_2$}
\Vertex(84,40){2}
\Vertex(80,20){2}
\Text(75,20)[r]{$\bar{q}$}
\SetColor{Red}
\ArrowLine(80,200)(84,181)
\ArrowLine(84,179)(88,161)
\ArrowLine(88,159)(92,141)
\ArrowLine(92,139)(94,121)
\ArrowLine(94,119)(94,101)
\ArrowLine(94,99)(92,81)
\ArrowLine(92,79)(88,61)
\ArrowLine(88,59)(84,41)
\ArrowLine(84,39)(80,21)
\SetColor{Black}
\LongArrow(120,110)(150,110)
\Vertex(200,200){2}
\Text(195,200)[r]{$q$}
\Vertex(204,180){2}
\Vertex(208,160){2}
\Text(213,160)[l]{$x_1$}
\Vertex(212,140){2}
\Text(217,140)[l]{$y_1$}
\Vertex(214,120){2}
\Vertex(214,100){2}
\Vertex(212,80){2}
\Text(217,80)[l]{$x_2$}
\Vertex(208,60){2}
\Text(213,60)[l]{$y_2$}
\Vertex(204,40){2}
\Vertex(200,20){2}
\Text(195,20)[r]{$\bar{q}$}
\SetColor{Red}
\ArrowLine(200,199)(204,181)
\ArrowLine(204,179)(208,161)
\Curve{(185,110)(186,115)(187,120)(190,130)(195,140)(201,150)(208,159)}
\Curve{(185,110)(186,105)(187,100)(190,90)(195,80)(201,70)(208,61)}
\LongArrow(185,111)(185,109)
\SetColor{Blue}
\ArrowLine(212,139)(214,121)
\ArrowLine(214,119)(214,101)
\ArrowLine(214,99)(212,81)
\Curve{(200,110)(201,106)(202,100)(205,90)(212,81)}
\Curve{(200,110)(201,114)(202,120)(205,130)(212,139)}
\LongArrow(200,109)(200,111)
\SetColor{Red}
\ArrowLine(208,59)(204,41)
\ArrowLine(204,39)(200,21)
\end{picture}
}}
\end{center}
\caption{\label{swing} Schematic picture of a colour recoupling, 
or dipole swing. The two dipoles $(\pmb{x}_1,\pmb{y}_1)$ and 
$(\pmb{x}_2,\pmb{y}_2)$ are transformed into two new dipoles 
$(\pmb{x}_1,\pmb{y}_2)$ and $(\pmb{x}_2,\pmb{y}_1)$ after a 
recoupling of the colour flow. The initial chain of dipoles 
is replaced by a new chain stretching between the original 
$q\bar{q}$ pair, red colour, and a loop of dipoles, blue colour.}
\end{figure}

We also want to emphasize that in our 
formalism the dipoles have a specified direction, going from colour to 
anticolour, and the cascade forms a connected and directed chain (see 
figure \ref{swing}). In this picture one cannot 
simply take two arbitrary dipoles and merge them into one 
dipole (as this would not lead to an allowed colour structure), 
as opposed to the effective picture described above. 
Such an effective picture is adequate if one just wants to calculate 
the total inclusive cross section to leading order. However, 
since we demand energy-momentum conservation and have the ambition to 
include final state radiation and hadronization in future work, 
it is absolutely necessary to keep track of this orientation 
of the dipoles. 

Another problem related to the finite number of colour charges
is the overcompleteness of the dipole basis. Beyond large $N_c$ 
more complicated colour structures will appear, and the charges in the
cascade cannot be represented in a unique way in terms of dipoles. 
The nonleading corrections in $N_c$ make the colour structure of the
gluon cascade really non-trivial, and one loses the picture of a 
system of dipoles evolving through dipole splittings in a stochastic
process. However, it may  still be possible to find a working approximation
using only dipoles. If, for example, the charges 
of a quadrupole are well separated, then it is possible 
to view such a quadrupole as two dipoles formed by the nearby 
colour charges. Such an approximative scheme can be realized by introducing 
a $2\rightarrow 2$ transition vertex in the dipole evolution. 

We choose this $2\rightarrow 2$ vertex such that it favours
the formation of dipoles by nearby colour charges. Initially
we have two dipoles $(\pmb{x}_1,\pmb{y}_1)$ and $(\pmb{x}_2,\pmb{y}_2)$
which might then ``swing'' into two new dipoles,  $(\pmb{x}_1,\pmb{y}_2)$ 
and $(\pmb{x}_2,\pmb{y}_1)$, as shown in figure \ref{swing}. The 
weight for this process is chosen to be 
\begin{eqnarray}
\mathcal{W} = \frac{\lambda}{N_c^2-1} \frac{(\pmb{x}_1-\pmb{y}_1)^2(\pmb{x}_2-\pmb{y}_2)^2}
{(\pmb{x}_1-\pmb{y}_2)^2(\pmb{x}_2-\pmb{y}_1)^2}
\end{eqnarray}
where $\lambda$ is a phenomenological parameter. Note that, with this 
choice of $\mathcal{W}$, the total weight for a dipole chain is 
$\propto \prod_i \frac{1}{r_i^2}$ where $r_i$ is the size of the 
``remaining dipoles'' in the cascade. Strictly speaking, if 
seen as a way to approximate more complicated colour structures, this 
process should happen instantly, rather than being proportional to 
$dY$. However, the value we choose for the phenomenological parameter, 
$\lambda =1$, turn out be quite interesting because increasing it 
does not change the cross section \cite{Avsar:2006jy}, so that the $2\rightarrow 2$ 
process can indeed be seen as happening instantly (for details see \cite{Avsar:2006jy}).  
Actually this process could also be seen as the result of a gluon 
exchange between $(\pmb{x}_1,\pmb{y}_1)$ and $(\pmb{x}_2,\pmb{y}_2)$, 
which would also result in the same colour structure. 

\section{Results}

In this section we present our results for $\gamma^*p$ and $pp$ scattering,
which we obtain from our MC. For now, we only calculate total
cross sections and all calculations are performed using a running 
coupling constant $\alpha_s$. In $\gamma^*p$ scattering the $\gamma^*$ 
fluctuates into a $q\bar{q}$ pair, which initiates the dipole cascade. 
This process can be described using perturbative methods and the 
results for the $\gamma^*$ longitudinal and transverse wave functions, 
$|\psi_L (z,r)|^2$ and $|\psi_T (z,r)|^2$ respectively, are well 
known. These wave functions give the probability for the $\gamma^*$
to fluctuate into a $q\bar{q}$ pair, separated by transverse distance
$\pmb{r}$ and where the quark carries a fraction $z$ of the $\gamma^*$
longitudinal momentum. 

\begin{figure}
\begin{center}
\scalebox{0.6}{\mbox{
\begin{picture}(100,50)(0,5)
\Vertex(50,50){2}
\Vertex(20,10){2}
\Vertex(80,10){2}
\SetColor{Red}
\Line(50,50)(20,10)
\SetColor{Blue}
\Line(20,10)(80,10)
\SetColor{Green}
\Line(80,10)(50,50)
\end{picture}
}}
\end{center}
\caption{\label{proton}The $\Delta$ shaped topology for the proton.}
\end{figure}
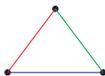

The proton end of the cascade is more complicated.
The easiest way to picture the proton would be to use a toy model 
which simply consists of a collection of uncorrelated dipoles. However, 
such a model fails to describe the energy dependence, and the
impact parameter profile (which is not black enough at $\pmb{b}=0$ 
and has a too long tail for large $\pmb{b}$), of the $pp$ cross 
section and it is seen that the dipoles inside the proton need to 
be more tightly correlated. Our main assumption is that a proton at
rest mainly consists of its three valence quarks. It is then natural 
to think that these quarks are the endpoints of three dipoles. 
Actually, one can try to calculate the emission probability of a 
soft gluon from these quarks. It turns out that they indeed form 
a system of three dipoles\footnote{With the emission probability suppressed by a factor 2, 
since we have 3 dipoles formed by 3 quarks and not 3 gluons.}\cite{Praszalowicz:1997nf}
so that the topology of the system equals that of a triangle, see figure \ref{proton}. 
The problem is however that further emissions do not factorize, even 
in the large $N_c$ limit\footnote{I would like to thank M. Praszalowicz
for pointing this out.}, as they do in the onium case. Nevertheless, 
as an approximation, we use the $\Delta$ topology for the initial system 
in the proton and we see that this works very well. 
\begin{figure}
  \begin{center}
    \includegraphics[angle=270, scale=0.5]{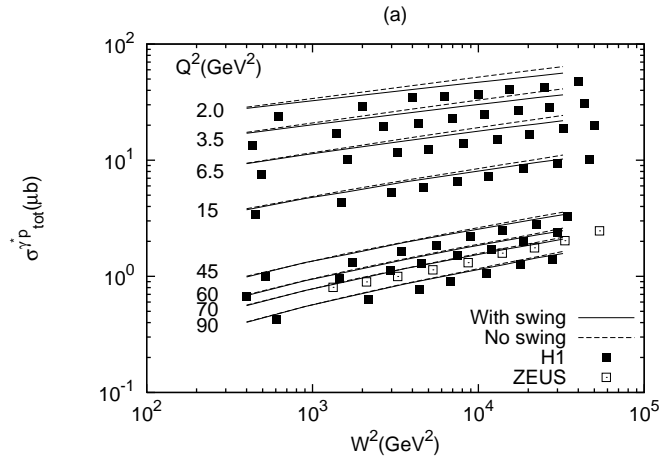}
    \caption{\label{gptot}The $\gamma^*p$ total cross section shown for 
      different $Q^2$. The solid lines include the dipole swing 
      while the dashed lines are without the dipole swing. $W$ denotes
      the cms energy. Data points are taken from \cite{Adloff:2000qk, Breitweg:2000yn}.}
  \end{center}
\end{figure}

For the proton model described above we choose the initial dipole 
sizes to be distributed as Gaussians with average size $\sim 3.1$GeV$^{-1}$ as 
this value agrees best with data. Obviously confinement must 
somehow suppress the formation of too large dipoles, and it is natural
to choose a maximum allowed dipole size. We choose this value to be 
the same as the average size of the initial dipoles in the proton as 
this corresponds to the nonperturbative input of the problem. Each 
new emission, giving a dipole of size $r$, is then allowed with 
a probability exp$(-r^2/r_{max}^2)$ where $r_{max}=3.1$GeV$^{-1}$. 

In figure \ref{gptot} we show the results for the $\gamma^*p$ total cross section. As we 
can see the results are in quite good agreement with data except for the fact that 
the normalization is around 10-15\% too high. We also see that the effects of the dipole
swing are quite small, visible for $Q^2 \lesssim$ 15 GeV$^2$. 
\begin{figure}
  \begin{center}
    \includegraphics[angle=270, scale=0.5]{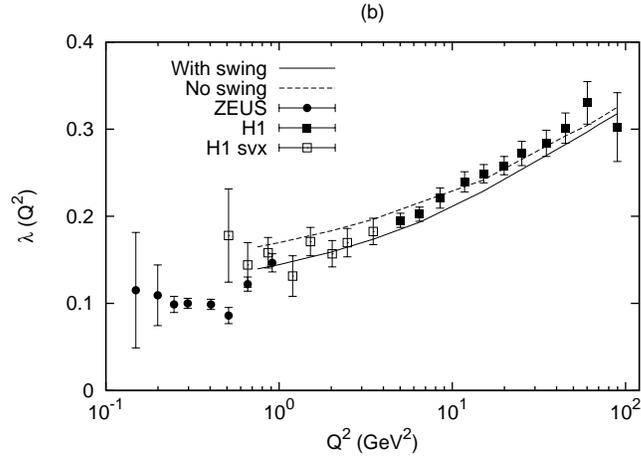}
    \caption{\label{effslope}The effective slope measured at different
      $Q^2$. The solid line is our result including the dipole swing
      while the dashed line is without. Filled circles are
      data from ZEUS\cite{Breitweg:2000yn} while
      filled\cite{Adloff:2000qk} and open\cite{Petrukhin:2004zy} squares
      are data from H1.}
  \end{center}
\end{figure}
In figure \ref{effslope} we show the results for the logarithmic slope   
$\lambda_{\mathrm{eff}}=d (\log \sigma)/d (\log 1/x)$. 
We see that there is a good agreement
with data for all points lying in the interval 
1GeV$^2 \lesssim Q^2 \lesssim$100 GeV$^2$. Here the slope 
is determined within the same energy interval from which the 
experimental points are determined. 
\begin{figure}
  \begin{center}
  \includegraphics[angle=270, scale=0.65]{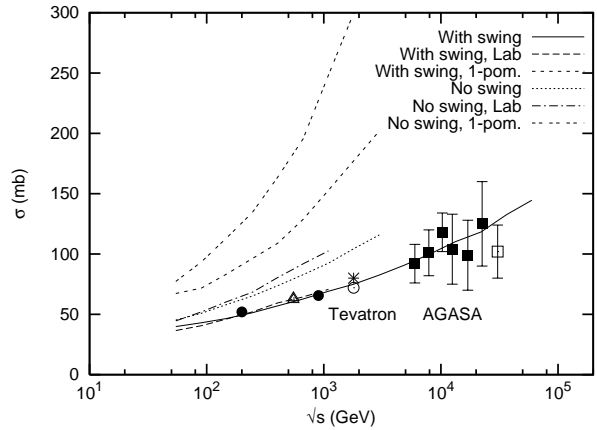}
  \caption{\label{pptotal}The total cross section for $pp$ scattering as 
    a function of the cms energy $\sqrt{s}$. Here results are shown 
    for evolution with and without the dipole swing mechanism. The results 
    for the one pomeron cross sections are also shown. Also shown are the 
    results obtained in the ``lab'' frame where one of the protons is almost at rest.}
    \end{center}
\end{figure}

Figure \ref{pptotal} shows the results for the $pp$ total 
cross section. Here it is seen that the dipole swing has a
rather large effect, as expected. Here we also show the results
in the one pomeron approximation and one can see the large effects
from unitarisation. These results are calculated 
in the center of mass frame where the colliding protons share 
the energy equally. In the figure we also show the result obtained 
in the ``lab'' frame where one of the protons carries almost all 
avaliable energy while the other one is essentially at rest. Due to the 
fact that the Monte Carlo simulation becomes very inefficient in 
such a frame (since the energetic proton has to be boosted to 
quite high rapidities) we have evaulated $\sigma_{tot}$ only 
up to $\sqrt{s} \sim 1$TeV. Although the result is not exactly frame
independent we see that the frame dependence
is reduced, and now very small. It is our intention to try to find a 
formalism where frame independence is explicit. It also turns out that the impact 
parameter profile of the cross section agrees well with data \cite{Avsar:2006jy}. 

I would like to thank the organizers and especially M. Praszalowicz for 
giving me the opportunity to come to the school and present my work.

\bibliographystyle{utcaps} 
\bibliography{refs2,refs}

\end{document}